\documentclass[%
 reprint,
 amsmath,amssymb,
 aps,
]{revtex4-1}

\usepackage{graphicx,float}
\usepackage{dcolumn}
\usepackage{bm}
\usepackage{subfigure}
\usepackage{braket}
\usepackage{xcolor}
\usepackage{tabularx}
\usepackage{multirow}

\newcommand{\comment}[1]{{}}

\begin{document}

\preprint{APS/123-QED}

\title{Quantifying Unknown Quantum Entanglement via a Hybrid Quantum-Classical Machine Learning Framework}

 \author{Xiaodie Lin$^{1}$}
 \author{Zhenyu Chen$^{1}$}
 \author{Zhaohui Wei$^{2,3,}$}\email{Email: weizhaohui@gmail.com}
 \affiliation{$^{1}$Institute for Interdisciplinary Information Sciences, Tsinghua University, Beijing 100084, China\\$^{2}$Yau Mathematical Sciences Center, Tsinghua University, Beijing 100084, China\\$^{3}$Yanqi Lake Beijing Institute of Mathematical Sciences and Applications, 101407, China}

\begin{abstract}

Quantifying unknown quantum entanglement experimentally is a difficult task, but also becomes more and more necessary because of the fast development of quantum engineering. Machine learning provides practical solutions to this fundamental problem, where one has to train a proper machine learning model to predict entanglement measures of unknown quantum states based on experimentally measurable data, say moments or correlation data produced by local measurements. In this paper, we compare the performance of these two different machine learning approaches systematically. Particularly, we first show that the approach based on moments enjoys a remarkable advantage over that based on correlation data, though the cost of measuring moments is much higher. Next, since correlation data is much easier to obtain experimentally, we try to better its performance by proposing a hybrid quantum-classical machine learning framework for this problem, where the key is to train optimal local measurements to generate more informative correlation data. Our numerical simulations show that the new framework brings us comparable performance with the approach based on moments to quantify unknown entanglement. Our work implies that it is already practical to fulfill such tasks on near-term quantum devices.

\end{abstract}

\maketitle

\section{Introduction}\label{introduction}

Quantum entanglement is a {crucial} resource for many quantum schemes and quantum protocols, such as {quantum superdense coding \cite{bennett1992communication}, quantum teleportation \cite{bennett1993teleporting} and quantum cryptography \cite{PhysRevLett.67.661}.} As a result, numerous measures have been raised to quantify the amount of entanglement contained in quantum states {\cite{horodecki2009quantum,amico2008entanglement}}. For bipartite pure states, entanglement measure is uniquely defined by the von Neumann entropy of subsystems. However, the landscape is far more complex for bipartite mixed states, where many important questions on entanglement quantifications have not been answered. An even more complicated case is the quantifications of multipartite quantum entanglement, for which many different measures have been proposed but most of them are very hard to calculate. It can be said that quantum entanglement has not been understood well theoretically.

Meanwhile, in recent years many subareas of quantum engineering are under fast development, and because of this, detecting and even quantifying \emph{unknown entanglement} involved in quantum experiments is becoming more and more realistic and necessary, though this is also a difficult task to fulfill. {For example, in a future factory that produces maximally entangled states by entanglement distillations, an efficient technique that can roughly quantify unknown entanglement will be helpful to select high quality raw materials.} We stress that unknown entanglement means that we do not have any prior knowledge of target quantum states, and therefore we have to characterize them and particularly quantify their entanglement by performing quantum measurements.

Existing quantification methods for unknown entanglement can be mainly divided into {four} branches. First, quantum tomography is the most popular approach adopted by quantum experimentalists when the size of target quantum states is small \cite{chuang1997prescription,poyatos1997complete}, where reconstructing quantum states allows us to look into the underlying entanglement. However, quantum tomography requires exponential cost, which is unbearable in high dimensional quantum systems. Second, recently a new technique of quantifying entanglement, say estimating the R\'enyi entropies, has been proposed, which performs random measurements on quantum states and then analyzes the outcome statistics \cite{van2012measuring,brydges2019probing}. Nevertheless, the cost of this technique is very high, {where a large number of measurement settings are necessary.} Third, device-independent protocols have been proposed to lower bound various entanglement measures \cite{moroder2013device,wei2021analytic,linli2021quantifying}. These device-independent methods quantify entanglement exclusively from the observed measurement statistics on subsystems, thus independent of any assumptions on the interested quantum systems. In the realm of noisy intermediate scale quantum (NISQ) era, device-independent protocols are attractive due to their efficiency and reliability. However, such device-independent protocols usually have limited applications in practical as they can provide nontrivial results only when the observed quantum nonlocality is very strong. {Fourth}, in addition to these analytical methods, machine learning has also been utilized to provide practical solutions to quantify entanglement experimentally \cite{gray2018machine,lin2021quantifying,roik2022entanglement}. In such methods, different experimentally accessible data on quantum states is collected and fed into machine learning models such that the mappings from experimental data to target entanglement measures are learned, by which one can predict the entanglement measures of unknown quantum states. In the current paper, we will focus on the {fourth} kind of methods.

In fact, according to the sorts of involved experimental data, there exist already two different approaches to apply machine learning onto quantifying unknown quantum entanglement experimentally \cite{gray2018machine,lin2021quantifying}. Specifically, in Ref.\cite{gray2018machine} the target entanglement measure is negativity, and for this the {moment data of partially transposed states} is fed into machine learning models as data features, which is usually very costly to obtain \cite{zhou2020single,elben2020mixed}. In Ref.\cite{lin2021quantifying}, correlation data serves as data features to quantify unknown entanglement, which is relatively convenient to prepare as one only needs to measure subsystems of target quantum states with a small set of local measurements chosen beforehand. 

In this paper, we first show that several entanglement measures, like the relative entropy of entanglement, can also be quantified accurately using machine learning models based on the original moments of quantum state $\rho$ defined as $\mu_m(\rho)={\rm Tr}\left(\rho^m\right)$. Then we compare this new approach with the one in Ref.\cite{lin2021quantifying}, and show that it can beat the latter easily in performance, though usually moment data are much harder to extract than correlation data. For example, a large number of measurement devices are already needed even if only $\mu_2(\rho)$ is measured \cite{vermersch2018unitary,brydges2019probing} . 


Meanwhile, since correlation data comes from a small set of local measurements and is much easier to collect, it will be nice if we can somehow improve the performance of the machine learning approach based on correlation data. Note that in the above comparisons, correlation data is generated by {a} certain fixed set of local measurements for all training and test quantum states. Therefore, a possible way to improve the performance is to choose better local measurements for correlation data generations. For this, we propose a hybrid quantum-classical {machine learning framework} to quantify unknown entanglement based on correlation data, where optimal local measurements are trained to generate correlation data. Our numerical simulations show that the new framework allows the correlation method to achieve a comparable performance with the machine learning approach based on moments in quantifying unknown entanglement {task}. Since the cost to produce correlation data is reasonable, our results show that it is already practical to quantify unknown entanglement on near-term quantum devices.


\section{Settings and Entanglement Measures}


Consider a bipartite state $\rho$ shared by separated two parties, Alice and Bob. Alice (respectively, Bob) has a set of measurement devices labeled by $X$ (respectively, $Y$) to measure her (his) subsystem, and the possible measurement outcomes are labeled by $A$ (respectively, $B$). After repeating the measurement many times, Alice and Bob calculate the joint conditional probabilities $p(ab|xy)$, which indicates the probability of obtaining outcomes $(a,b)\in A\times B$ upon selecting measurement settings $(x,y)\in X\times Y$. Suppose $\{M_x^a\}$ is the operators for the quantum measurement performed by Alice's measurement device $x\in X$, where $a\in A$, and analogously for $\{N_y^b\}$, then it holds that
\begin{eqnarray}\label{Correlation}
p(ab|xy)={\rm Tr}\left[\left(M_x^a\otimes N_y^b\right)\rho\right].
\end{eqnarray}
A \emph{correlation} $p=[p(ab|xy)]$ is a vector contains all the joint conditional probabilities of form $p(ab|xy)$.

We now turn to \emph{moments} of quantum states, which {are} defined as {\cite{ekert2002direct,horodecki2003measuring}}
\begin{eqnarray}\label{Moments}
\mu_m(\rho)={\rm Tr}\left(\rho^m\right).
\end{eqnarray}
Clearly, $\mu_1(\rho)={\rm Tr}(\rho)=1$ and $\mu_2(\rho)$ is the \emph{purity} of $\rho$. {Experimentally}, $\mu_m(\rho)$ can be measured directly by performing joint measurements on $m$ copies of the same state $\rho$ {\cite{ekert2002direct}}, while this operation is very hard with current quantum technologies, especially when $m$ is large. {To overcome this difficulty, techniques that can estimate $\mu_m(\rho)$ on single-copy states have also been developed \cite{van2012measuring}.}

A {complete} supervised machine learning system contains three ingredients: data features, data labels and a machine learning model {\cite{mohri2018foundations,shalev2014understanding}}. The training data features with correct labels are fed into the machine learning model, where the parameters {contained in the learning model} are trained properly such that the resulting system can predict labels of unknown test data precisely. In our problem, correlation data or moments serve as data features of quantum states, data labels are the values of target entanglement measures, and the machine learning model is supposed to learn an unknown nonlinear relationship between data features and data labels. According to the no free lunch theorem \cite{585893}, 
a small target error rate implies that we need a large set of representative training data. However, it is well-known that for bipartite mixed quantum states and multipartite quantum states, most entanglement measures are extremely hard to calculate, which means that it is hard for us to prepare correct labels for training data {sets}. According to the computation hardness and the importance of entanglement measures, in this paper we choose the coherent information and the relative entropy of entanglement as our target measures to quantify unknown quantum entanglement.

Coherent information is a fundamental quantity that measures the capability of transition of quantum information \cite{schumacher1996quantum,lloyd1997capacity}. For an arbitrary bipartite quantum state $\rho\in\mathcal{H}_A\otimes \mathcal{H}_B$, its coherent information is defined as
\begin{eqnarray}\label{IC}
I_C(\rho)=S(\rho_A)-S(\rho),
\end{eqnarray}
where $S(\rho)$ is the von Neumann entropy of $\rho$ and $\rho_A={\rm Tr}_B(\rho)$ is the subsystem in $\mathcal{H}_A$. A crucial property {of} the coherent information is that for any bipartite $\rho$, it holds that {\cite{cornelio2011entanglement}}
\begin{eqnarray}
E_F(\rho)\geq E_D(\rho)\geq I_C(\rho),
\end{eqnarray}
where $E_F(\rho)$ and $E_D(\rho)$ are two most important measures of entanglement, the entanglement of formation and the entanglement of distillation respectively. Therefore, a good estimation of $I_C(\rho)$ implies that we obtain a very nontrivial piece of information on the amount of entanglement for $\rho$. Furthermore, it is not hard to see that the coherent information is very easy to calculate, which means that {if we pick it as} the data label, we can generate a large amount of training or test data at low costs.

Another quantity we will utilize is the relative entropy of entanglement \cite{vedral2002role}, defined as
\begin{eqnarray}\label{RelativeEntropy}
E_R(\rho)=\min_{\sigma\in {\rm SEP}}S(\rho\|\sigma)=\min_{\sigma\in {\rm SEP}}{\rm Tr}(\rho\log \rho-\rho\log\sigma),
\end{eqnarray}
where ${\rm SEP}$ denotes the set of all separable states. Relative entropy of entanglement has a good geometric interpretation, as $E_R(\rho)$ measures certain ``distance" between $\rho$ and the set of separable states.
Based on a machine learning model called \emph{active learning}, we can numerically compute the relative entropy of entanglement for some particular quantum states with high accuracy \cite{hou2020upper}. At the same time, it turns out that the relative entropy of entanglement satisfies that {\cite{vedral1998entanglement}}
\begin{eqnarray}\label{EM_Relationship}
E_F(\rho)\ge E_R(\rho)\ge E_D(\rho).
\end{eqnarray}
Therefore, similar {to} $I_C(\rho)$, estimating $E_R(\rho)$ of an unknown quantum state $\rho$ also helps us to obtain nontrivial information {of} $E_F(\rho)$ and $E_D(\rho)$.

\section{Comparisons between machine learning models based on orrelation data and moments}\label{Comparison}

In this section, we first present machine learning models that take correlation data and moments as data features respectively, then we compare their performance in predicting both the coherent information and the relative entropy of entanglement systematically. For convenience, we call the former the correlation method, and the latter the moment method.

\subsection{Quantifying the Coherent Information}\label{CoherentInformation}

We begin with quantifying the coherent information of arbitrary 3-dimensional bipartite quantum states $\rho\in\mathcal{H}^3\otimes\mathcal{H}^3$. To generate more representative training data, we try to evenly sample quantum states according to the distribution of their coherent information. However, due to the low efficiency of sampling quantum states with coherent information less than -1.5, in fact we only sample quantum states with coherent information ranging from -1.5 to $\log_23$. Specifically, we divide this range into 31 intervals of size 0.1, and in each of these intervals we sample 1,291 quantum states randomly. Eventually, totally 40,021 states are randomly sampled to compose the set of quantum states {for training}.

After sampling the training quantum states, we generate the corresponding training data features for the correlation method, which is achieved by performing the local measurements that maximize the violation of the {Collins-Gisin-Linden-Masser-Popescu (CGLMP)} inequality on the training quantum states {\cite{collins2002bell,zohren2008maximal}}, and then record the outcome statistics. More concretely, Alice's measurement $A_k$ can be characterized by the eigenvectors
\begin{eqnarray}\label{CGLMP_Measurement1}
|r\rangle_{A_k}=\frac{1}{\sqrt{d}}\sum_{q=0}^{d-1}{\rm exp}\left(\frac{2\pi i}{d}q(r-\alpha_k)\right)|q\rangle_A,
\end{eqnarray}
and Bob's measurement $B_l$ can be characterized by the eigenvectors
\begin{eqnarray}\label{CGLMP_Measurement2}
|r\rangle_{B_l}=\frac{1}{\sqrt{d}}\sum_{q=0}^{d-1}{\rm exp}\left(-\frac{2\pi i}{d}q(r-\beta_l)\right)|q\rangle_B,
\end{eqnarray}
where $0\leq r\leq d-1$, $1\leq k,l\leq N$, $\alpha_k=(k-1/2)/N,\beta_l=l/N$, and $N=|X|=|Y|$ {\cite{barrett2006maximally}}. In the current task, we let $d=3$ and $N=2$. That is, both Alice and Bob have $2$ different measurement devices. 

Meanwhile, to apply the moment method, we generate two different sets of training data, which contain different orders of moments. Specifically, one set has $\{\mu_2(\rho_A),\mu_2(\rho)\}$ as data features, and the other has $\{\mu_2(\rho_A),\mu_2(\rho),\mu_3(\rho_A),\mu_3(\rho)\}$ as data features.

To test the performance of the above three cases (one for the correlation method and two for the moment method), we {evenly sample around 2,000 quantum states}, and then produce the corresponding test correlation data or moments {of these states}.

\begin{figure}[!ht]
    \centering
    \includegraphics[width=0.45\textwidth]{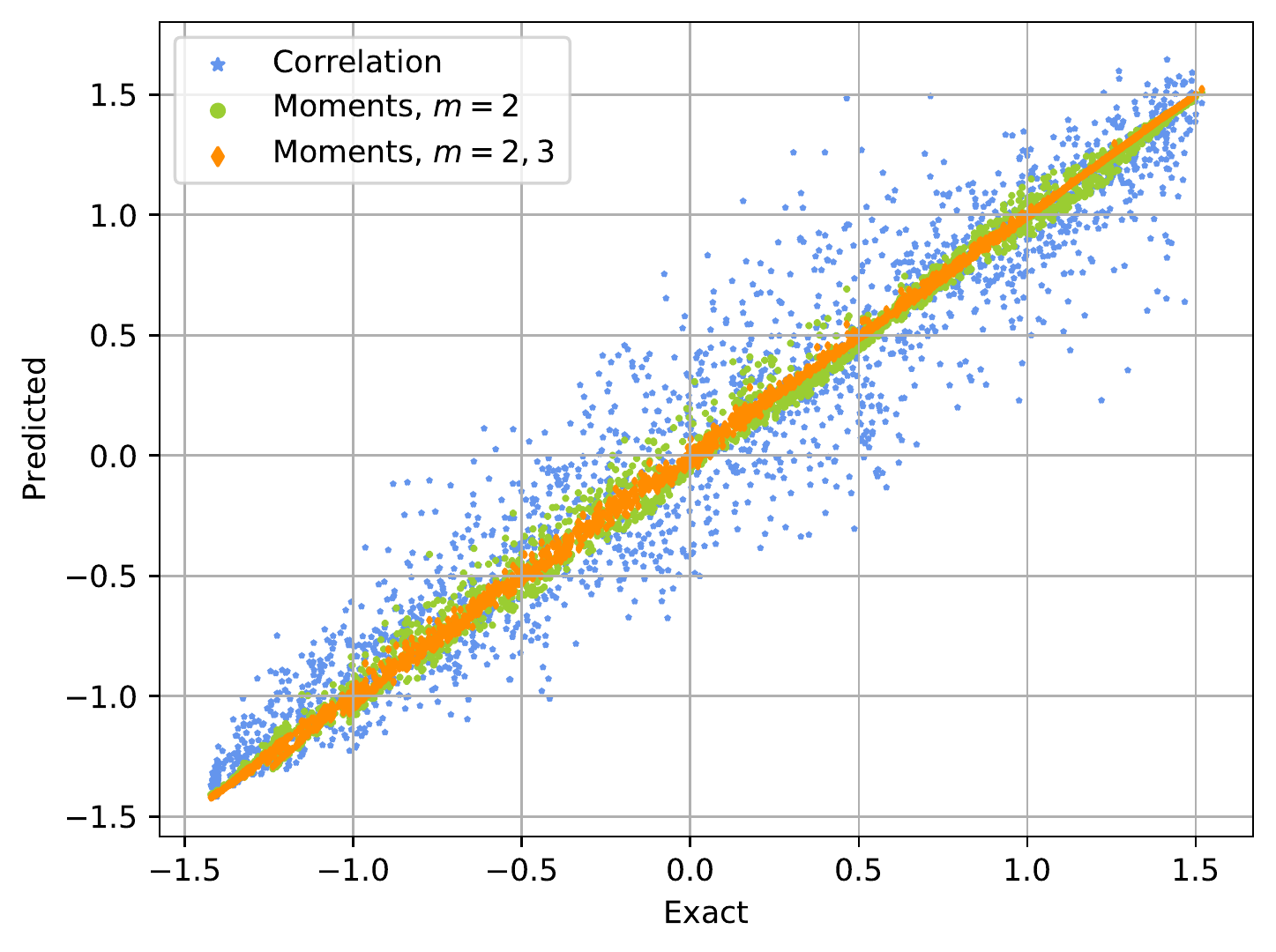}
    \caption{The neural network predictions for $I_C(\rho)$. The blue stars represent the prediction values of the correlation method and the corresponding MSE is 0.0582. The green dots and orange diamonds represent the prediction values of the moment method with $m=2$ and $m=2,3$, whose MSE are $0.0035$ and $0.0004$, respectively.}
    \label{fig:result_General}
\end{figure}

During the training stage, each training data set is fed into a 4-hidden-layer fully connected neural network (FNN) with 400, 200, 100 and 50 neurons in each layer, respectively. After training, we test its performance with the corresponding {test} set sampled above. The results of all the three cases are shown in Fig.\ref{fig:result_General}. As we can see, the moment method that utilizes only the second order of moment already beats the correlation method in this task, whose corresponding mean squared errors (MSEs) are 0.0582 and 0.0035, respectively. In some sense this is not surprising since the number of measurement devices required by the moment method is much larger than that of the correlation method, and thus the moment data is more informative.

\begin{table*}[htb] \scriptsize
    \centering
    \caption {The structure and configuration details of the convolutional neural network (see Refs.\cite{lecun1998gradient,albawi2017understanding} for introductions to convolutional neural networks). For each `$\cdot/\cdot$', the former parameter represents for $d=5$ and the latter parameter represents for $d=8,10$.}
    \setlength{\tabcolsep}{4.5mm}
   {
    \begin{tabular}{lccccccl}
        \hline
        \hline
        Layers &Type&Neurons&Filters&Kernel size&Strides&Pool size\\
        \hline
        0-1 & Convolution2D & (None, 9/18, 9/18, 32) & 32  & $2\times2$/$3\times 3$ & $1\times 1$ & -\\
        1-2 & Max-pooling2D & (None, 8/16, 8/16, 32) & - & - & $1\times 1$ & $2\times 2$/$3\times 3$\\
        2-3 & Convolution2D & (None, 7/14, 7/14, 64) & 64  & $2\times 2$/$3\times 3$ & $1\times 1$ & -\\
        3-4 & Max-pooling2D & (None, 6/12, 6/12, 64) & - & - & $1\times 1$ & $2\times 2$/$3\times 3$\\
        4-5 & Convolution2D & (None, 5/10, 5/10, 64)  & 64  & $2\times 2$/$3\times 3$ & $1\times 1$ & -\\
        5-6 & Fully-connected & (None, 64) & -  & - & - & -\\
        6-7 & Fully-connected & (None, 32) & -  & - & - & -\\
        7-8 & Fully-connected & (None, 1) & -  & - & - & -\\
        \hline
        \hline
    \end{tabular}}
    \label{table:CNN}
\end{table*}

In addition, if we strengthen the moment method by also factoring in the third order of moment, apparent further improvements can be observed, where the MSE decreases to 0.0004, implying an excellent performance. 
As a result, compared with the correlation method, the moment method enjoys {a} remarkable advantage in coherent information quantification tasks.

We now increase the dimension of target quantum states $\rho\in\mathcal{H}^d\otimes\mathcal{H}^d$ to $d=5,8,10$, and still aim to quantify the coherent information. In this case, the quantum state sampling procedure is similar {to} the dimension 3 case. Firstly, around 40,000 quantum states are evenly sampled for training according to the distribution of coherent information, and then around 2,000 quantum states are sampled for test similarly. The generations of data features for these quantum states are exactly the same as the previous task.

The training models for the moment method keep unchanged, {i.e.,} a 4-hidden-layer FNN with 400, 200, 100 and 50 neurons in each {hidden} layer is adopted. While for the correlation method, since convolutional neural networks (CNNs) behave better than FNNs when dimension increases, in the current task we utilize CNNs to predict the coherent information instead of FNNs. Table \ref{table:CNN} describes the structure and configuration details of our CNN model.

Applying the well-traind models on the test data sets we choose, the experimental results for each case are listed in Table \ref{table:result_diffdim}.
\begin{table}[H] \scriptsize
\caption{The MSEs of predicting the coherent information for different methods and dimensions.}
\setlength{\tabcolsep}{3mm}{
\begin{tabular}{lcccccc}
    \hline
    \hline
    Dimension & $d=3$ & $d=5$ & $d=8$ & $d=10$\\
    \hline
    Correlation, $N=2$ & 0.0582 & 0.0362 & 0.0212 & 0.0146\\
    Moments, $m=2$ & 0.0035 & 0.0034 & 0.0025 & 0.0026\\
    Moments, $m=2,3$ & 0.0004 & 0.0010 & 0.0011 & 0.0010\\
    \hline
    \hline
\end{tabular}}
\label{table:result_diffdim}
\end{table}
Similar {to} the dimension 3 case, there still exists an obvious gap between the performance of the correlation method and that of the moment method. Furthermore, it is interesting to see that the improvements of introducing the third order of moment ($m=2,3$) over only using the second order ($m=2$) decreases as {the} quantum dimension goes up. Therefore, considering the experimental {difficulty} of measuring moments, it is a good choice to set $m=2$ when predicting coherent information for high dimensional quantum states.

\subsection{Quantifying the Relative Entropy of Entanglement}

Recall that the concept of the relative entropy of entanglement is defined as an optimization problem over the set of separable states. Despite being convex, the set of separable states is still very hard to fully characterize {\cite{lu2018separability,doherty2002distinguishing}}. In Ref.\cite{hou2020upper}, a technique to upper bound the relative entropy of entanglement was proposed based on active learning. Even though this method only provides upper bounds, Ref.\cite{hou2020upper} demonstrated that these upper bounds are quite tight for many quantum states, such as Werner states, isotropic states and random {bipartite} quantum states with low {dimensions}. However, it should be stressed that the method introduced in Ref.\cite{hou2020upper} requires full descriptions of quantum states, {i.e., the density matrix of the state}, while our mission is quantifying unknown entanglement based on experimentally measuring quantities. In our {task}, we only utilize the active learning method to provide labels for our data sets, due to its high accuracy.

As usual, to generate representative training data, we would like to sample quantum states with evenly distributed relative entropy of entanglement. However, when the target values of relative entropy of entanglement are high, the sampling efficiency is very low. In addition, to provide correct labels for the sampling quantum states, we have to calculate their relative entropy of entanglement using the active learning method introduced above, which is very costly. Both of these two facts make it challenging for us to generate proper training and test data.

Here to prepare training data we focus on quantum states with the form
\begin{eqnarray}\label{REE_states}
\rho=(1-\epsilon)\rho_0+\epsilon|\psi_+\rangle\langle\psi_+|,
\end{eqnarray}
where $\epsilon\in[0,1]$, $\rho_0$ is {a random }quantum state in $\mathcal{H}^d\otimes\mathcal{H}^d$ and $|\psi_+\rangle=\frac{1}{\sqrt{d}}\sum_i|ii\rangle$ is the maximally entangled state. For each case of the dimension $d=2,3,4$, 3,000 states are sampled by randomly selecting $\epsilon$ and $\rho_0$. Together with 1,000 random separable states, totally 4,000 states are sampled to serve as the training quantum states.

We next generate training data based on {the sampled} quantum states. For the correlation method, we prepare the training data by measuring the quantum states via the local measurements given in Eqs.(\ref{CGLMP_Measurement1}) and (\ref{CGLMP_Measurement2}), where for each party the number of measurement devices $N$ is fixed as 2. For the moment method, the moments $\{\mu_m(\rho_A),\mu_m(\rho_B),\mu_m(\rho)\}$ are chosen as data features. Again, two sets of training data with $m=2$ and $m=2,3$ are generated, in order to compare the power of moments of different orders.

As mentioned, the labels of the training data are provided by the active learning method. Meanwhile, the machine learning models for all the three cases are the same, which is a 4-hidden-layer FNN with 400, 200, 100 and 50 neurons in each layer. 

\begin{figure}[H]
    \centering
    \includegraphics[width=0.43\textwidth]{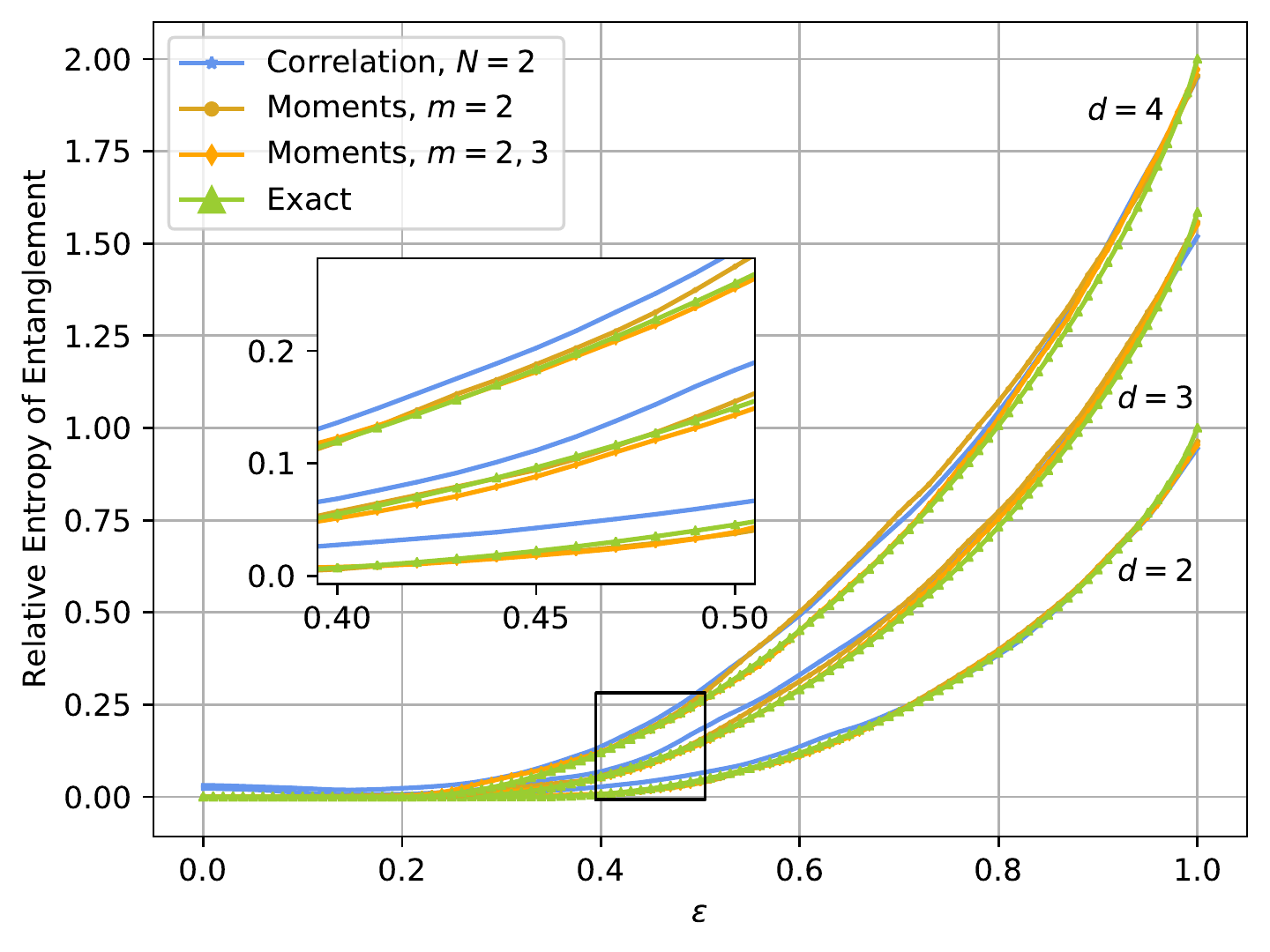}
    \caption{The neural network predictions for $E_R(\rho_{d_\epsilon}^{me})$. The blue stars, grown dots and orange diamonds represent the prediction values of the correlation method, the moment method with $m=2$ and the moment method with $m=2,3$, respectively. The exact values are represented by the green triangles. All the corresponding MSEs are listed in the first part of Table \ref{table:result_bipartite}.}\label{fig:result_ree_bipartite}
\end{figure}

After training, to estimate the performance of these models, we first run them on isotropic states, whose parameterized form is given by
\begin{eqnarray}\label{istropic}
\rho_{d_\epsilon}^{me}=\frac{(1-\epsilon)}{d^2}I_{d^2}+\epsilon|\psi_+\rangle\langle\psi_+|,
\end{eqnarray}
where $\epsilon\in[0,1]$. Given an isotropic state, its relative entropy of entanglement can be analytically calculated \cite{wei2008relative}. We make the comparisons between the results given by all the three cases and the exact results for $d=2,3,4$. The results are depicted in Fig.\ref{fig:result_ree_bipartite}, and the corresponding MSEs are shown in the first part of Table \ref{table:result_bipartite}.

\begin{table}[H] \scriptsize
\caption{The MSEs of predicting the relative entropy of entanglement for different methods and dimensions.}
\setlength{\tabcolsep}{1mm}{
\begin{tabular}{llccccc}
    \hline
    \hline
    \multicolumn{2}{l}{Dimension} & $d=2$ & $d=3$ & $d=4$\\
    \hline
    \multirow{3}{*}{Isotropic}
    & Correlation, $N=2$ & $1.71\times10^{-4}$ & $4.14\times10^{-4}$ & $6.45\times10^{-4}$ \\
    & Moments, $m=2$ & $3.57\times10^{-5}$ & $2.57\times10^{-3}$ & $9.08\times10^{-4}$ \\
    & Moments, $m=2,3$ & $4.88\times10^{-5}$ & $6.92\times10^{-5}$ & $1.37\times10^{-4}$ \\
    \hline
    \multirow{3}{*}{General} & Correlation, $N=2$ & $2.99\times10^{-3}$ & $6.90\times10^{-3}$ & $7.59\times10^{-3}$ \\
    & Moments, $m=2$ & $2.26\times10^{-4}$ & $9.26\times10^{-4}$ & $1.45\times10^{-3}$ \\
    & Moments, $m=2,3$ & $2.01\times10^{-4}$ & $5.75\times10^{-4}$ & $9.12\times10^{-4}$ \\
    \hline
    \hline
\end{tabular}}
\label{table:result_bipartite}
\end{table}

As illustrated, all the three cases (one for the correlation method and two for the moment method) predict the relative entropy of entanglement of isotropic states correctly with high precision. Particularly, the results given by the active learning method, {whose MESs are in the order of $10^{-7}\sim10^{-5}$, match the analytical results given by Ref.\cite{wei2008relative} accurately.}

Since the active learning method provides us a reliable way to calculate relative entropy of entanglement, we now apply it to provide labels for random quantum states, which allows us to test our models on more general quantum states, not just isotropic ones.

Specifically, for each dimension, we sample 300 quantum states admit the form Eq.(\ref{REE_states}), and combined with 100 random separable states, these 400 quantum states serve as test quantum states. The generations of data features for different methods are the same as before. Then we apply our trained FNNs to predict the relative entropy of entanglement of these quantum states, the corresponding MSEs are listed in the second part of Table \ref{table:result_bipartite}. 

It can be seen that in this task the overall behaviors of the MSEs of the correlation method and the moment method are similar to those in the coherent information prediction tasks. Actually in this task the performance of the correlation method is even better, but a stable advantage of the moment method can still be observed.

\section{Hybrid Quantum-Classical Framework assisted Correlation Method}\label{hybrid}

In Subsection.\ref{Comparison}, we fixed local measurements as the ones that achieve the maximal violation of the CGLMP inequality, and fixed the number of measurement devices for each party to be $N=2$. Due to these two constraints, the power of the correlation method may be underestimated. Hence, a natural question is: Can we improve the performance of the correlation method by relaxing the constraints, say, enlarging the number of measurement devices $N$ or changing the measurement devices?


\subsection{More Measurement Devices}\label{largerN}

Intuitively, enlarging the set of available measurement devices can probably provide more information about target quantum states, and therefore may improve the performance of the correlation method. To check whether this idea works, we demonstrate it by setting $N=3,4$ in the coherent information prediction tasks.

For each quantum dimension $d=3,5,8,10$, the sampled training and test quantum states, and the mathematical structures of machine learning models keep the same as before. The only difference is that now the training and {test} data sets are composed {of} measurement outcome statistics involving $N=3,4$ measurement devices, rather than 2. It turns out that very limited improvements are achieved by this change in predicting coherent information. Table \ref{table:result_largerN} lists the corresponding MSEs.

\begin{table}[h] \scriptsize
\caption{The MSEs of predicting the coherent information with fixed measurement devices.}
\setlength{\tabcolsep}{3mm}{
\begin{tabular}{lcccccc}
    \hline
    \hline
    Dimension & $d=3$ & $d=5$ & $d=8$ & $d=10$\\
    \hline
    CGLMP, $N=3$ & 0.0584 & 0.0356 & 0.0180 & 0.0170\\
    CGLMP, $N=4$ & 0.0540 & 0.0364 & 0.0183 & 0.0138\\
    \hline
    \hline
\end{tabular}}
\label{table:result_largerN}
\end{table}

The limited improvements given by increasing the number of measurement devices {are} unexpected, because more measurement devices should have revealed more information. A possible reason is that though the number of the measurement devices we have utilized were increased, they are still of the form in Eqs.(\ref{CGLMP_Measurement1}) and (\ref{CGLMP_Measurement2}). Therefore, to improve the performance of the correlation method further, we need to find out whether this form is optimal or not.

\subsection{Learnable Measurement Devices}\label{learnable}

\begin{figure*}[!ht]
    \centering
    \includegraphics[width=0.9\textwidth]{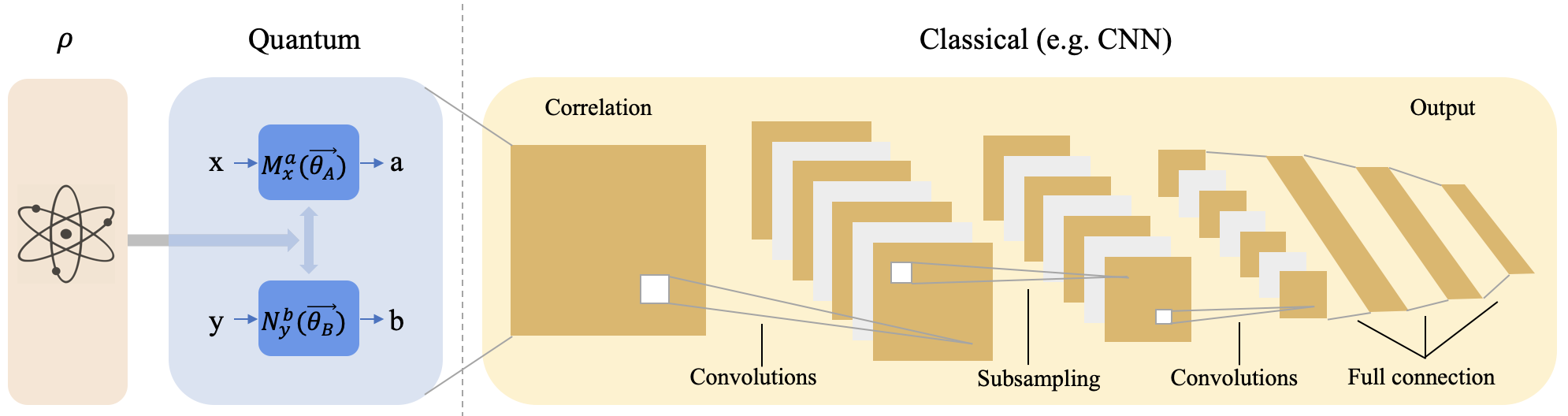}
    \caption{The hybrid quantum-classical framework for the correlation method. (1) The input quantum state $\rho$, for which the full information is required. (2) The generation of correlation, where the local measurements $M_x^a$ and $N_y^b$ contain trainable variables. (3) The classical training part, which can be, for example, a CNN.}
    \label{fig:hybrid_framework}
\end{figure*}

Looking back at all our previous machine learning models that have been discussed, we will see that all of them are classical models that deal with pure classical information, where all data features of involved quantum states are about correlation data or moments, which are essentially classical. However, we now need to choose {better} local measurements to generate more informative correlation data, which means that we have new quantum structures to learn. For this purpose, below we introduce a hybrid quantum-classical framework for our machine learning tasks.


In fact, a series of hybrid quantum-classical algorithms have been proposed \cite{peruzzo2014variational,mitarai2018quantum,mcclean2016theory,mari2020transfer,liu2021hybrid}, where the concept of parameterized quantum {circuits (PQCs)} is widely used to optimize a target function by iteratively tuning parameters contained in underlying quantum circuits. While in our hybrid quantum-classical framework, the tuning target is quantum measurements. Since a general quantum measurement can be realized by first performing a unitary operation and then measuring in the computational basis, what we are aiming at is essentially learning $2N$ such unitary operators.

Our hybrid quantum-classical framework works as {follows}. First, since we want to find the best local measurements for our tasks, the entries of observables $\{M_x\}$ and $\{N_y\}$ are now regarded as trainable variables and will be updated repeatedly, which is the quantum part of our hybrid {quantum-classical} machine learning framework. Second, after choosing the training quantum states, we need to generate their training data features by measuring the parameterized observables $\{M_x\}$ and $\{N_y\}$, and then collecting the outcome statistics. Along with the correct labels, these data features will be fed into the classical learning part, which is the same {as the} previous models we have discussed. {Fig.\ref{fig:hybrid_framework} illustrates the whole framework.} It is worth mentioning that according to our previous experience, different permutations of the measurement devices can result in quite different prediction performance, especially when the training model is a CNN. Therefore, introducing trainable measurements will bring us the optimized permutation automatically. 

\begin{figure}[!ht]
    \centering
    \includegraphics[width=0.45\textwidth]{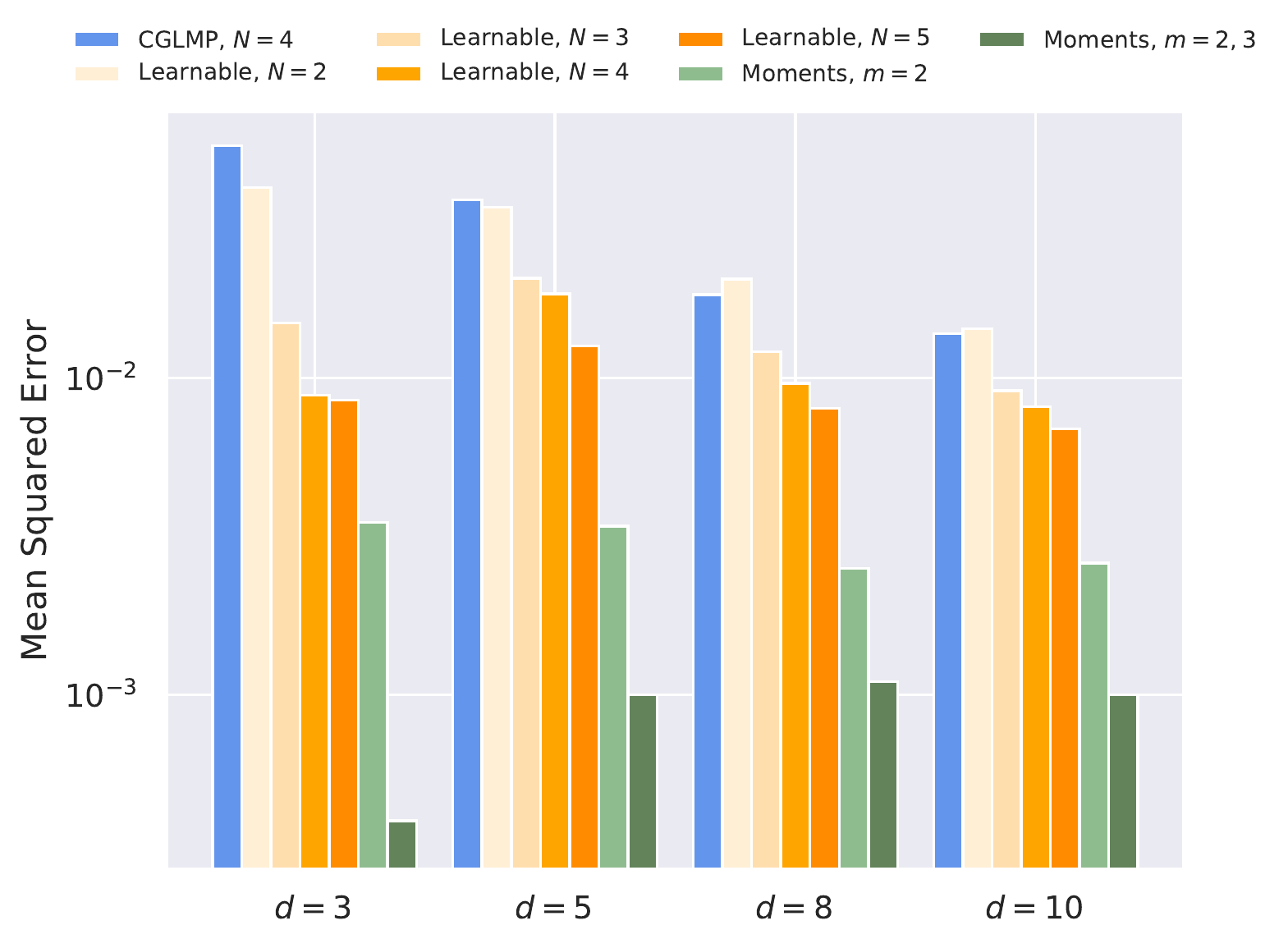}
    \caption{Comparisons between the correlation method using the CGLMP measurements, learnable measurements and the moment method. The MSEs of the correlation method using learnable measurements decrease as $N$ increases. Particularly, when $N=3$ an apparent improvement is achieved. When {$N=4$}, the hybrid model even has the same order of MSE as the moment method with $m=2,3$ {in high dimensions}. The values of each MSE are listed in Table \ref{table:result_diffM}.}
    \label{fig:result_Learnable}
\end{figure}

We apply the hybrid quantum-classical {machine learning} framework {to} the coherent information prediction tasks for the cases of $N=2,3,4,5$. The sampled training and test quantum states, and the mathematical structure of the classical machine learning part remain the same as Section.\ref{CoherentInformation}, where the dimension $d=3,5,8,10$. The experimental results are listed in Fig.\ref{fig:result_Learnable} and Table \ref{table:result_diffM}.

\begin{table}[h] \scriptsize
\caption{The MSEs of predicting the coherent information with learnable measurement devices.}
\setlength{\tabcolsep}{3mm}{
\begin{tabular}{lcccccc}
    \hline
    \hline
    Dimension & $d=3$ & $d=5$ & $d=8$ & $d=10$\\
    \hline
    Learnable, $N=2$ & 0.0398 & 0.0345 & 0.0205 & 0.0143\\
    Learnable, $N=3$ & 0.0149 & 0.0206 & 0.0121 & 0.0091\\
    Learnable, $N=4$ & 0.0088 & 0.0184 & 0.0096 & 0.0081\\
    Learnable, $N=5$ & 0.0085 & 0.0126 & 0.0080 & 0.0069\\
    \hline
    \hline
\end{tabular}}
\label{table:result_diffM}
\end{table}

Comparing the case $N=2$ with the old results in Table \ref{table:result_diffdim}, it can be seen that almost no improvement is achieved for {each dimension except $d=3$}, which indicates that in generating statistics data the local measurements in Eqs.(\ref{CGLMP_Measurement1}) and (\ref{CGLMP_Measurement2}) are {almost optimal for these cases.} However, once we enlarge the number of possible measurement devices by only one, that is $N=3$, the situation becomes totally different, {where the corresponding MSEs are 0.0206, 0.0121 and 0.0091 for dimensions $5,8$ and $10$, respectively. Compared with the original version of the correlation method, $43.09\%,42.92\%$ and $37.67\%$ improvements are achieved for dimensions $5,8$ and $10$, respectively.} 

Furthermore, if we increase the number of measurement devices further, the improvements are even more remarkable. As illustrated in Fig.\ref{fig:result_Learnable}, when $d = 3$, our hybrid model with {$N=4$} has the same order of MSE as the moment method with $m = 2$, and when {$d = 8,10$}, it is even comparable with that of the moment method with $m = 2,3$. {In fact, it turns out that even if we restrict Alice and Bob to perform the same set of local measurements, the obtained MSEs differ very little in high dimensions, e.g., $d=5,8,10$. Therefore, a half number of the variables in the quantum part can be reduced to achieve a similar performance, hence facilitating the training efficiency.}

Recall that though the moment method has very good performance, the experimental cost of measuring moments is much higher than that of generating correlation data. Therefore, our result clearly implies that our hybrid quantum-classical framework of machine learning largely overcomes this difficulty, and achieves similar performance. This also shows that it is already practical to quantify unknown entanglement on current NISQ devices.

\section{Conclusion}

Quantifying unknown entanglement experimentally is a very challenging task today. However, considering the profound importance of quantum entanglement and the fast developments of quantum industries, fulfilling this kind of tasks will probably become a daily routine in the future. As a result, finding realistic and economical solutions for this problem from the viewpoints of engineering is extremely necessary, and a typical one of this kind is using machine learning methods.

In this paper, we {focus} on two known machine learning approaches for quantifying unknown entanglement experimentally, one using moments as data features, and the other using correlation data. We systematically compare their performance in predicting the coherent information and the relative entropy of entanglement. According to our results, the moment method behaves much better than the correlation method, which is consistent with the intuition. In fact, as mentioned, each moment is a quantity related to eigenvalues of the corresponding density matrix, and all the moments together in principle can pin down the essential information of the density matrix. Hence, the moment method naturally enjoys  remarkable {advantage} in quantifying entanglement. However, estimating moments is much harder than correlation data.

This motivates us to improve the performance of the correlation method in quantifying entanglement. For this, we propose a hybrid quantum-classical framework of machine learning models to generate more informative correlation data. Specifically, we studied two possible directions to achieve this, where one directly enlarges the number of local measurements fixed beforehand to produce more informative correlation data, and the other adds a new quantum module for the machine learning model to search for better local measurements utilized in correlation data generations, and at the same time increases their number. {It} turns out that the former direction only has little improvement, while the latter is much better, and it even achieves comparable performance with the moment method. Therefore, it is fair to say the hybrid quantum-classical {machine learning} framework we propose has very good performance in quantifying unknown entanglement experimentally. And we hope that in {the} future such protocols can be deployed on NISQ devices.


\begin{acknowledgements}
This work was supported by the National Key R\&D Program of China, Grants No. 2018YFA0306703, 2021YFE0113100, and the National Natural Science Foundation of China, Grant No. 61832015.
\end{acknowledgements}

\bibliography{main}

\begin{thebibliography}{42}%
\makeatletter
\providecommand \@ifxundefined [1]{%
 \@ifx{#1\undefined}
}%
\providecommand \@ifnum [1]{%
 \ifnum #1\expandafter \@firstoftwo
 \else \expandafter \@secondoftwo
 \fi
}%
\providecommand \@ifx [1]{%
 \ifx #1\expandafter \@firstoftwo
 \else \expandafter \@secondoftwo
 \fi
}%
\providecommand \natexlab [1]{#1}%
\providecommand \enquote  [1]{``#1''}%
\providecommand \bibnamefont  [1]{#1}%
\providecommand \bibfnamefont [1]{#1}%
\providecommand \citenamefont [1]{#1}%
\providecommand \href@noop [0]{\@secondoftwo}%
\providecommand \href [0]{\begingroup \@sanitize@url \@href}%
\providecommand \@href[1]{\@@startlink{#1}\@@href}%
\providecommand \@@href[1]{\endgroup#1\@@endlink}%
\providecommand \@sanitize@url [0]{\catcode `\\12\catcode `\$12\catcode
  `\&12\catcode `\#12\catcode `\^12\catcode `\_12\catcode `\%12\relax}%
\providecommand \@@startlink[1]{}%
\providecommand \@@endlink[0]{}%
\providecommand \url  [0]{\begingroup\@sanitize@url \@url }%
\providecommand \@url [1]{\endgroup\@href {#1}{\urlprefix }}%
\providecommand \urlprefix  [0]{URL }%
\providecommand \Eprint [0]{\href }%
\providecommand \doibase [0]{http://dx.doi.org/}%
\providecommand \selectlanguage [0]{\@gobble}%
\providecommand \bibinfo  [0]{\@secondoftwo}%
\providecommand \bibfield  [0]{\@secondoftwo}%
\providecommand \translation [1]{[#1]}%
\providecommand \BibitemOpen [0]{}%
\providecommand \bibitemStop [0]{}%
\providecommand \bibitemNoStop [0]{.\EOS\space}%
\providecommand \EOS [0]{\spacefactor3000\relax}%
\providecommand \BibitemShut  [1]{\csname bibitem#1\endcsname}%
\let\auto@bib@innerbib\@empty
\bibitem [{\citenamefont {Bennett}\ and\ \citenamefont
  {Wiesner}(1992)}]{bennett1992communication}%
  \BibitemOpen
  \bibfield  {author} {\bibinfo {author} {\bibfnamefont {C.~H.}\ \bibnamefont
  {Bennett}}\ and\ \bibinfo {author} {\bibfnamefont {S.~J.}\ \bibnamefont
  {Wiesner}},\ }\href@noop {} {\bibfield  {journal} {\bibinfo  {journal}
  {Physical review letters}\ }\textbf {\bibinfo {volume} {69}},\ \bibinfo
  {pages} {2881} (\bibinfo {year} {1992})}\BibitemShut {NoStop}%
\bibitem [{\citenamefont {Bennett}\ \emph {et~al.}(1993)\citenamefont
  {Bennett}, \citenamefont {Brassard}, \citenamefont {Cr{\'e}peau},
  \citenamefont {Jozsa}, \citenamefont {Peres},\ and\ \citenamefont
  {Wootters}}]{bennett1993teleporting}%
  \BibitemOpen
  \bibfield  {author} {\bibinfo {author} {\bibfnamefont {C.~H.}\ \bibnamefont
  {Bennett}}, \bibinfo {author} {\bibfnamefont {G.}~\bibnamefont {Brassard}},
  \bibinfo {author} {\bibfnamefont {C.}~\bibnamefont {Cr{\'e}peau}}, \bibinfo
  {author} {\bibfnamefont {R.}~\bibnamefont {Jozsa}}, \bibinfo {author}
  {\bibfnamefont {A.}~\bibnamefont {Peres}}, \ and\ \bibinfo {author}
  {\bibfnamefont {W.~K.}\ \bibnamefont {Wootters}},\ }\href@noop {} {\bibfield
  {journal} {\bibinfo  {journal} {Physical review letters}\ }\textbf {\bibinfo
  {volume} {70}},\ \bibinfo {pages} {1895} (\bibinfo {year}
  {1993})}\BibitemShut {NoStop}%
\bibitem [{\citenamefont {Ekert}(1991)}]{PhysRevLett.67.661}%
  \BibitemOpen
  \bibfield  {author} {\bibinfo {author} {\bibfnamefont {A.~K.}\ \bibnamefont
  {Ekert}},\ }\href {\doibase 10.1103/PhysRevLett.67.661} {\bibfield  {journal}
  {\bibinfo  {journal} {Phys. Rev. Lett.}\ }\textbf {\bibinfo {volume} {67}},\
  \bibinfo {pages} {661} (\bibinfo {year} {1991})}\BibitemShut {NoStop}%
\bibitem [{\citenamefont {Horodecki}\ \emph {et~al.}(2009)\citenamefont
  {Horodecki}, \citenamefont {Horodecki}, \citenamefont {Horodecki},\ and\
  \citenamefont {Horodecki}}]{horodecki2009quantum}%
  \BibitemOpen
  \bibfield  {author} {\bibinfo {author} {\bibfnamefont {R.}~\bibnamefont
  {Horodecki}}, \bibinfo {author} {\bibfnamefont {P.}~\bibnamefont
  {Horodecki}}, \bibinfo {author} {\bibfnamefont {M.}~\bibnamefont
  {Horodecki}}, \ and\ \bibinfo {author} {\bibfnamefont {K.}~\bibnamefont
  {Horodecki}},\ }\href@noop {} {\bibfield  {journal} {\bibinfo  {journal}
  {Reviews of modern physics}\ }\textbf {\bibinfo {volume} {81}},\ \bibinfo
  {pages} {865} (\bibinfo {year} {2009})}\BibitemShut {NoStop}%
\bibitem [{\citenamefont {Amico}\ \emph {et~al.}(2008)\citenamefont {Amico},
  \citenamefont {Fazio}, \citenamefont {Osterloh},\ and\ \citenamefont
  {Vedral}}]{amico2008entanglement}%
  \BibitemOpen
  \bibfield  {author} {\bibinfo {author} {\bibfnamefont {L.}~\bibnamefont
  {Amico}}, \bibinfo {author} {\bibfnamefont {R.}~\bibnamefont {Fazio}},
  \bibinfo {author} {\bibfnamefont {A.}~\bibnamefont {Osterloh}}, \ and\
  \bibinfo {author} {\bibfnamefont {V.}~\bibnamefont {Vedral}},\ }\href@noop {}
  {\bibfield  {journal} {\bibinfo  {journal} {Reviews of modern physics}\
  }\textbf {\bibinfo {volume} {80}},\ \bibinfo {pages} {517} (\bibinfo {year}
  {2008})}\BibitemShut {NoStop}%
\bibitem [{\citenamefont {Chuang}\ and\ \citenamefont
  {Nielsen}(1997)}]{chuang1997prescription}%
  \BibitemOpen
  \bibfield  {author} {\bibinfo {author} {\bibfnamefont {I.~L.}\ \bibnamefont
  {Chuang}}\ and\ \bibinfo {author} {\bibfnamefont {M.~A.}\ \bibnamefont
  {Nielsen}},\ }\href@noop {} {\bibfield  {journal} {\bibinfo  {journal}
  {Journal of Modern Optics}\ }\textbf {\bibinfo {volume} {44}},\ \bibinfo
  {pages} {2455} (\bibinfo {year} {1997})}\BibitemShut {NoStop}%
\bibitem [{\citenamefont {Poyatos}\ \emph {et~al.}(1997)\citenamefont
  {Poyatos}, \citenamefont {Cirac},\ and\ \citenamefont
  {Zoller}}]{poyatos1997complete}%
  \BibitemOpen
  \bibfield  {author} {\bibinfo {author} {\bibfnamefont {J.}~\bibnamefont
  {Poyatos}}, \bibinfo {author} {\bibfnamefont {J.~I.}\ \bibnamefont {Cirac}},
  \ and\ \bibinfo {author} {\bibfnamefont {P.}~\bibnamefont {Zoller}},\
  }\href@noop {} {\bibfield  {journal} {\bibinfo  {journal} {Physical Review
  Letters}\ }\textbf {\bibinfo {volume} {78}},\ \bibinfo {pages} {390}
  (\bibinfo {year} {1997})}\BibitemShut {NoStop}%
\bibitem [{\citenamefont {Van~Enk}\ and\ \citenamefont
  {Beenakker}(2012)}]{van2012measuring}%
  \BibitemOpen
  \bibfield  {author} {\bibinfo {author} {\bibfnamefont {S.}~\bibnamefont
  {Van~Enk}}\ and\ \bibinfo {author} {\bibfnamefont {C.}~\bibnamefont
  {Beenakker}},\ }\href@noop {} {\bibfield  {journal} {\bibinfo  {journal}
  {Physical review letters}\ }\textbf {\bibinfo {volume} {108}},\ \bibinfo
  {pages} {110503} (\bibinfo {year} {2012})}\BibitemShut {NoStop}%
\bibitem [{\citenamefont {Brydges}\ \emph {et~al.}(2019)\citenamefont
  {Brydges}, \citenamefont {Elben}, \citenamefont {Jurcevic}, \citenamefont
  {Vermersch}, \citenamefont {Maier}, \citenamefont {Lanyon}, \citenamefont
  {Zoller}, \citenamefont {Blatt},\ and\ \citenamefont
  {Roos}}]{brydges2019probing}%
  \BibitemOpen
  \bibfield  {author} {\bibinfo {author} {\bibfnamefont {T.}~\bibnamefont
  {Brydges}}, \bibinfo {author} {\bibfnamefont {A.}~\bibnamefont {Elben}},
  \bibinfo {author} {\bibfnamefont {P.}~\bibnamefont {Jurcevic}}, \bibinfo
  {author} {\bibfnamefont {B.}~\bibnamefont {Vermersch}}, \bibinfo {author}
  {\bibfnamefont {C.}~\bibnamefont {Maier}}, \bibinfo {author} {\bibfnamefont
  {B.~P.}\ \bibnamefont {Lanyon}}, \bibinfo {author} {\bibfnamefont
  {P.}~\bibnamefont {Zoller}}, \bibinfo {author} {\bibfnamefont
  {R.}~\bibnamefont {Blatt}}, \ and\ \bibinfo {author} {\bibfnamefont {C.~F.}\
  \bibnamefont {Roos}},\ }\href@noop {} {\bibfield  {journal} {\bibinfo
  {journal} {Science}\ }\textbf {\bibinfo {volume} {364}},\ \bibinfo {pages}
  {260} (\bibinfo {year} {2019})}\BibitemShut {NoStop}%
\bibitem [{\citenamefont {Moroder}\ \emph {et~al.}(2013)\citenamefont
  {Moroder}, \citenamefont {Bancal}, \citenamefont {Liang}, \citenamefont
  {Hofmann},\ and\ \citenamefont {G{\"u}hne}}]{moroder2013device}%
  \BibitemOpen
  \bibfield  {author} {\bibinfo {author} {\bibfnamefont {T.}~\bibnamefont
  {Moroder}}, \bibinfo {author} {\bibfnamefont {J.-D.}\ \bibnamefont {Bancal}},
  \bibinfo {author} {\bibfnamefont {Y.-C.}\ \bibnamefont {Liang}}, \bibinfo
  {author} {\bibfnamefont {M.}~\bibnamefont {Hofmann}}, \ and\ \bibinfo
  {author} {\bibfnamefont {O.}~\bibnamefont {G{\"u}hne}},\ }\href@noop {}
  {\bibfield  {journal} {\bibinfo  {journal} {Physical review letters}\
  }\textbf {\bibinfo {volume} {111}},\ \bibinfo {pages} {030501} (\bibinfo
  {year} {2013})}\BibitemShut {NoStop}%
\bibitem [{\citenamefont {Wei}\ and\ \citenamefont
  {Lin}(2021)}]{wei2021analytic}%
  \BibitemOpen
  \bibfield  {author} {\bibinfo {author} {\bibfnamefont {Z.}~\bibnamefont
  {Wei}}\ and\ \bibinfo {author} {\bibfnamefont {L.}~\bibnamefont {Lin}},\
  }\href@noop {} {\bibfield  {journal} {\bibinfo  {journal} {Physical Review
  A}\ }\textbf {\bibinfo {volume} {103}},\ \bibinfo {pages} {032215} (\bibinfo
  {year} {2021})}\BibitemShut {NoStop}%
\bibitem [{\citenamefont {Lin}\ and\ \citenamefont
  {Wei}(2021)}]{linli2021quantifying}%
  \BibitemOpen
  \bibfield  {author} {\bibinfo {author} {\bibfnamefont {L.}~\bibnamefont
  {Lin}}\ and\ \bibinfo {author} {\bibfnamefont {Z.}~\bibnamefont {Wei}},\
  }\href@noop {} {\bibfield  {journal} {\bibinfo  {journal} {Physical Review
  A}\ }\textbf {\bibinfo {volume} {104}},\ \bibinfo {pages} {062433} (\bibinfo
  {year} {2021})}\BibitemShut {NoStop}%
\bibitem [{\citenamefont {Gray}\ \emph {et~al.}(2018)\citenamefont {Gray},
  \citenamefont {Banchi}, \citenamefont {Bayat},\ and\ \citenamefont
  {Bose}}]{gray2018machine}%
  \BibitemOpen
  \bibfield  {author} {\bibinfo {author} {\bibfnamefont {J.}~\bibnamefont
  {Gray}}, \bibinfo {author} {\bibfnamefont {L.}~\bibnamefont {Banchi}},
  \bibinfo {author} {\bibfnamefont {A.}~\bibnamefont {Bayat}}, \ and\ \bibinfo
  {author} {\bibfnamefont {S.}~\bibnamefont {Bose}},\ }\href@noop {} {\bibfield
   {journal} {\bibinfo  {journal} {Physical review letters}\ }\textbf {\bibinfo
  {volume} {121}},\ \bibinfo {pages} {150503} (\bibinfo {year}
  {2018})}\BibitemShut {NoStop}%
\bibitem [{\citenamefont {Lin}\ \emph {et~al.}(2021)\citenamefont {Lin},
  \citenamefont {Chen},\ and\ \citenamefont {Wei}}]{lin2021quantifying}%
  \BibitemOpen
  \bibfield  {author} {\bibinfo {author} {\bibfnamefont {X.}~\bibnamefont
  {Lin}}, \bibinfo {author} {\bibfnamefont {Z.}~\bibnamefont {Chen}}, \ and\
  \bibinfo {author} {\bibfnamefont {Z.}~\bibnamefont {Wei}},\ }\href@noop {}
  {\bibfield  {journal} {\bibinfo  {journal} {arXiv preprint arXiv:2104.12527}\
  } (\bibinfo {year} {2021})}\BibitemShut {NoStop}%
\bibitem [{\citenamefont {Roik}\ \emph {et~al.}(2022)\citenamefont {Roik},
  \citenamefont {Bartkiewicz}, \citenamefont {{\v{C}}ernoch},\ and\
  \citenamefont {Lemr}}]{roik2022entanglement}%
  \BibitemOpen
  \bibfield  {author} {\bibinfo {author} {\bibfnamefont {J.}~\bibnamefont
  {Roik}}, \bibinfo {author} {\bibfnamefont {K.}~\bibnamefont {Bartkiewicz}},
  \bibinfo {author} {\bibfnamefont {A.}~\bibnamefont {{\v{C}}ernoch}}, \ and\
  \bibinfo {author} {\bibfnamefont {K.}~\bibnamefont {Lemr}},\ }\href@noop {}
  {\bibfield  {journal} {\bibinfo  {journal} {arXiv preprint arXiv:2203.01607}\
  } (\bibinfo {year} {2022})}\BibitemShut {NoStop}%
\bibitem [{\citenamefont {Zhou}\ \emph {et~al.}(2020)\citenamefont {Zhou},
  \citenamefont {Zeng},\ and\ \citenamefont {Liu}}]{zhou2020single}%
  \BibitemOpen
  \bibfield  {author} {\bibinfo {author} {\bibfnamefont {Y.}~\bibnamefont
  {Zhou}}, \bibinfo {author} {\bibfnamefont {P.}~\bibnamefont {Zeng}}, \ and\
  \bibinfo {author} {\bibfnamefont {Z.}~\bibnamefont {Liu}},\ }\href@noop {}
  {\bibfield  {journal} {\bibinfo  {journal} {Physical Review Letters}\
  }\textbf {\bibinfo {volume} {125}},\ \bibinfo {pages} {200502} (\bibinfo
  {year} {2020})}\BibitemShut {NoStop}%
\bibitem [{\citenamefont {Elben}\ \emph {et~al.}(2020)\citenamefont {Elben},
  \citenamefont {Kueng}, \citenamefont {Huang}, \citenamefont {van Bijnen},
  \citenamefont {Kokail}, \citenamefont {Dalmonte}, \citenamefont {Calabrese},
  \citenamefont {Kraus}, \citenamefont {Preskill}, \citenamefont {Zoller} \emph
  {et~al.}}]{elben2020mixed}%
  \BibitemOpen
  \bibfield  {author} {\bibinfo {author} {\bibfnamefont {A.}~\bibnamefont
  {Elben}}, \bibinfo {author} {\bibfnamefont {R.}~\bibnamefont {Kueng}},
  \bibinfo {author} {\bibfnamefont {H.-Y.~R.}\ \bibnamefont {Huang}}, \bibinfo
  {author} {\bibfnamefont {R.}~\bibnamefont {van Bijnen}}, \bibinfo {author}
  {\bibfnamefont {C.}~\bibnamefont {Kokail}}, \bibinfo {author} {\bibfnamefont
  {M.}~\bibnamefont {Dalmonte}}, \bibinfo {author} {\bibfnamefont
  {P.}~\bibnamefont {Calabrese}}, \bibinfo {author} {\bibfnamefont
  {B.}~\bibnamefont {Kraus}}, \bibinfo {author} {\bibfnamefont
  {J.}~\bibnamefont {Preskill}}, \bibinfo {author} {\bibfnamefont
  {P.}~\bibnamefont {Zoller}},  \emph {et~al.},\ }\href@noop {} {\bibfield
  {journal} {\bibinfo  {journal} {Physical Review Letters}\ }\textbf {\bibinfo
  {volume} {125}},\ \bibinfo {pages} {200501} (\bibinfo {year}
  {2020})}\BibitemShut {NoStop}%
\bibitem [{\citenamefont {Vermersch}\ \emph {et~al.}(2018)\citenamefont
  {Vermersch}, \citenamefont {Elben}, \citenamefont {Dalmonte}, \citenamefont
  {Cirac},\ and\ \citenamefont {Zoller}}]{vermersch2018unitary}%
  \BibitemOpen
  \bibfield  {author} {\bibinfo {author} {\bibfnamefont {B.}~\bibnamefont
  {Vermersch}}, \bibinfo {author} {\bibfnamefont {A.}~\bibnamefont {Elben}},
  \bibinfo {author} {\bibfnamefont {M.}~\bibnamefont {Dalmonte}}, \bibinfo
  {author} {\bibfnamefont {J.~I.}\ \bibnamefont {Cirac}}, \ and\ \bibinfo
  {author} {\bibfnamefont {P.}~\bibnamefont {Zoller}},\ }\href@noop {}
  {\bibfield  {journal} {\bibinfo  {journal} {Physical Review A}\ }\textbf
  {\bibinfo {volume} {97}},\ \bibinfo {pages} {023604} (\bibinfo {year}
  {2018})}\BibitemShut {NoStop}%
\bibitem [{\citenamefont {Ekert}\ \emph {et~al.}(2002)\citenamefont {Ekert},
  \citenamefont {Alves}, \citenamefont {Oi}, \citenamefont {Horodecki},
  \citenamefont {Horodecki},\ and\ \citenamefont {Kwek}}]{ekert2002direct}%
  \BibitemOpen
  \bibfield  {author} {\bibinfo {author} {\bibfnamefont {A.~K.}\ \bibnamefont
  {Ekert}}, \bibinfo {author} {\bibfnamefont {C.~M.}\ \bibnamefont {Alves}},
  \bibinfo {author} {\bibfnamefont {D.~K.}\ \bibnamefont {Oi}}, \bibinfo
  {author} {\bibfnamefont {M.}~\bibnamefont {Horodecki}}, \bibinfo {author}
  {\bibfnamefont {P.}~\bibnamefont {Horodecki}}, \ and\ \bibinfo {author}
  {\bibfnamefont {L.~C.}\ \bibnamefont {Kwek}},\ }\href@noop {} {\bibfield
  {journal} {\bibinfo  {journal} {Physical review letters}\ }\textbf {\bibinfo
  {volume} {88}},\ \bibinfo {pages} {217901} (\bibinfo {year}
  {2002})}\BibitemShut {NoStop}%
\bibitem [{\citenamefont {Horodecki}(2003)}]{horodecki2003measuring}%
  \BibitemOpen
  \bibfield  {author} {\bibinfo {author} {\bibfnamefont {P.}~\bibnamefont
  {Horodecki}},\ }\href@noop {} {\bibfield  {journal} {\bibinfo  {journal}
  {Physical review letters}\ }\textbf {\bibinfo {volume} {90}},\ \bibinfo
  {pages} {167901} (\bibinfo {year} {2003})}\BibitemShut {NoStop}%
\bibitem [{\citenamefont {Mohri}\ \emph {et~al.}(2018)\citenamefont {Mohri},
  \citenamefont {Rostamizadeh},\ and\ \citenamefont
  {Talwalkar}}]{mohri2018foundations}%
  \BibitemOpen
  \bibfield  {author} {\bibinfo {author} {\bibfnamefont {M.}~\bibnamefont
  {Mohri}}, \bibinfo {author} {\bibfnamefont {A.}~\bibnamefont {Rostamizadeh}},
  \ and\ \bibinfo {author} {\bibfnamefont {A.}~\bibnamefont {Talwalkar}},\
  }\href@noop {} {\emph {\bibinfo {title} {Foundations of machine learning}}}\
  (\bibinfo  {publisher} {MIT press},\ \bibinfo {year} {2018})\BibitemShut
  {NoStop}%
\bibitem [{\citenamefont {Shalev-Shwartz}\ and\ \citenamefont
  {Ben-David}(2014)}]{shalev2014understanding}%
  \BibitemOpen
  \bibfield  {author} {\bibinfo {author} {\bibfnamefont {S.}~\bibnamefont
  {Shalev-Shwartz}}\ and\ \bibinfo {author} {\bibfnamefont {S.}~\bibnamefont
  {Ben-David}},\ }\href@noop {} {\emph {\bibinfo {title} {Understanding machine
  learning: From theory to algorithms}}}\ (\bibinfo  {publisher} {Cambridge
  university press},\ \bibinfo {year} {2014})\BibitemShut {NoStop}%
\bibitem [{\citenamefont {Wolpert}\ and\ \citenamefont
  {Macready}(1997)}]{585893}%
  \BibitemOpen
  \bibfield  {author} {\bibinfo {author} {\bibfnamefont {D.}~\bibnamefont
  {Wolpert}}\ and\ \bibinfo {author} {\bibfnamefont {W.}~\bibnamefont
  {Macready}},\ }\href {\doibase 10.1109/4235.585893} {\bibfield  {journal}
  {\bibinfo  {journal} {IEEE Transactions on Evolutionary Computation}\
  }\textbf {\bibinfo {volume} {1}},\ \bibinfo {pages} {67} (\bibinfo {year}
  {1997})}\BibitemShut {NoStop}%
\bibitem [{\citenamefont {Schumacher}\ and\ \citenamefont
  {Nielsen}(1996)}]{schumacher1996quantum}%
  \BibitemOpen
  \bibfield  {author} {\bibinfo {author} {\bibfnamefont {B.}~\bibnamefont
  {Schumacher}}\ and\ \bibinfo {author} {\bibfnamefont {M.~A.}\ \bibnamefont
  {Nielsen}},\ }\href@noop {} {\bibfield  {journal} {\bibinfo  {journal}
  {Physical Review A}\ }\textbf {\bibinfo {volume} {54}},\ \bibinfo {pages}
  {2629} (\bibinfo {year} {1996})}\BibitemShut {NoStop}%
\bibitem [{\citenamefont {Lloyd}(1997)}]{lloyd1997capacity}%
  \BibitemOpen
  \bibfield  {author} {\bibinfo {author} {\bibfnamefont {S.}~\bibnamefont
  {Lloyd}},\ }\href@noop {} {\bibfield  {journal} {\bibinfo  {journal}
  {Physical Review A}\ }\textbf {\bibinfo {volume} {55}},\ \bibinfo {pages}
  {1613} (\bibinfo {year} {1997})}\BibitemShut {NoStop}%
\bibitem [{\citenamefont {Cornelio}\ \emph {et~al.}(2011)\citenamefont
  {Cornelio}, \citenamefont {de~Oliveira},\ and\ \citenamefont
  {Fanchini}}]{cornelio2011entanglement}%
  \BibitemOpen
  \bibfield  {author} {\bibinfo {author} {\bibfnamefont {M.~F.}\ \bibnamefont
  {Cornelio}}, \bibinfo {author} {\bibfnamefont {M.~C.}\ \bibnamefont
  {de~Oliveira}}, \ and\ \bibinfo {author} {\bibfnamefont {F.~F.}\ \bibnamefont
  {Fanchini}},\ }\href@noop {} {\bibfield  {journal} {\bibinfo  {journal}
  {Physical review letters}\ }\textbf {\bibinfo {volume} {107}},\ \bibinfo
  {pages} {020502} (\bibinfo {year} {2011})}\BibitemShut {NoStop}%
\bibitem [{\citenamefont {Vedral}(2002)}]{vedral2002role}%
  \BibitemOpen
  \bibfield  {author} {\bibinfo {author} {\bibfnamefont {V.}~\bibnamefont
  {Vedral}},\ }\href@noop {} {\bibfield  {journal} {\bibinfo  {journal}
  {Reviews of Modern Physics}\ }\textbf {\bibinfo {volume} {74}},\ \bibinfo
  {pages} {197} (\bibinfo {year} {2002})}\BibitemShut {NoStop}%
\bibitem [{\citenamefont {Hou}\ \emph {et~al.}(2020)\citenamefont {Hou},
  \citenamefont {Cao}, \citenamefont {Zhou},\ and\ \citenamefont
  {Zeng}}]{hou2020upper}%
  \BibitemOpen
  \bibfield  {author} {\bibinfo {author} {\bibfnamefont {S.-Y.}\ \bibnamefont
  {Hou}}, \bibinfo {author} {\bibfnamefont {C.}~\bibnamefont {Cao}}, \bibinfo
  {author} {\bibfnamefont {D.}~\bibnamefont {Zhou}}, \ and\ \bibinfo {author}
  {\bibfnamefont {B.}~\bibnamefont {Zeng}},\ }\href@noop {} {\bibfield
  {journal} {\bibinfo  {journal} {Quantum Science and Technology}\ }\textbf
  {\bibinfo {volume} {5}},\ \bibinfo {pages} {045019} (\bibinfo {year}
  {2020})}\BibitemShut {NoStop}%
\bibitem [{\citenamefont {Vedral}\ and\ \citenamefont
  {Plenio}(1998)}]{vedral1998entanglement}%
  \BibitemOpen
  \bibfield  {author} {\bibinfo {author} {\bibfnamefont {V.}~\bibnamefont
  {Vedral}}\ and\ \bibinfo {author} {\bibfnamefont {M.~B.}\ \bibnamefont
  {Plenio}},\ }\href@noop {} {\bibfield  {journal} {\bibinfo  {journal}
  {Physical Review A}\ }\textbf {\bibinfo {volume} {57}},\ \bibinfo {pages}
  {1619} (\bibinfo {year} {1998})}\BibitemShut {NoStop}%
\bibitem [{\citenamefont {Collins}\ \emph {et~al.}(2002)\citenamefont
  {Collins}, \citenamefont {Gisin}, \citenamefont {Linden}, \citenamefont
  {Massar},\ and\ \citenamefont {Popescu}}]{collins2002bell}%
  \BibitemOpen
  \bibfield  {author} {\bibinfo {author} {\bibfnamefont {D.}~\bibnamefont
  {Collins}}, \bibinfo {author} {\bibfnamefont {N.}~\bibnamefont {Gisin}},
  \bibinfo {author} {\bibfnamefont {N.}~\bibnamefont {Linden}}, \bibinfo
  {author} {\bibfnamefont {S.}~\bibnamefont {Massar}}, \ and\ \bibinfo {author}
  {\bibfnamefont {S.}~\bibnamefont {Popescu}},\ }\href@noop {} {\bibfield
  {journal} {\bibinfo  {journal} {Physical review letters}\ }\textbf {\bibinfo
  {volume} {88}},\ \bibinfo {pages} {040404} (\bibinfo {year}
  {2002})}\BibitemShut {NoStop}%
\bibitem [{\citenamefont {Zohren}\ and\ \citenamefont
  {Gill}(2008)}]{zohren2008maximal}%
  \BibitemOpen
  \bibfield  {author} {\bibinfo {author} {\bibfnamefont {S.}~\bibnamefont
  {Zohren}}\ and\ \bibinfo {author} {\bibfnamefont {R.~D.}\ \bibnamefont
  {Gill}},\ }\href@noop {} {\bibfield  {journal} {\bibinfo  {journal} {Physical
  review letters}\ }\textbf {\bibinfo {volume} {100}},\ \bibinfo {pages}
  {120406} (\bibinfo {year} {2008})}\BibitemShut {NoStop}%
\bibitem [{\citenamefont {Barrett}\ \emph {et~al.}(2006)\citenamefont
  {Barrett}, \citenamefont {Kent},\ and\ \citenamefont
  {Pironio}}]{barrett2006maximally}%
  \BibitemOpen
  \bibfield  {author} {\bibinfo {author} {\bibfnamefont {J.}~\bibnamefont
  {Barrett}}, \bibinfo {author} {\bibfnamefont {A.}~\bibnamefont {Kent}}, \
  and\ \bibinfo {author} {\bibfnamefont {S.}~\bibnamefont {Pironio}},\
  }\href@noop {} {\bibfield  {journal} {\bibinfo  {journal} {Physical Review
  Letters}\ }\textbf {\bibinfo {volume} {97}},\ \bibinfo {pages} {170409}
  (\bibinfo {year} {2006})}\BibitemShut {NoStop}%
\bibitem [{\citenamefont {LeCun}\ \emph {et~al.}(1998)\citenamefont {LeCun},
  \citenamefont {Bottou}, \citenamefont {Bengio},\ and\ \citenamefont
  {Haffner}}]{lecun1998gradient}%
  \BibitemOpen
  \bibfield  {author} {\bibinfo {author} {\bibfnamefont {Y.}~\bibnamefont
  {LeCun}}, \bibinfo {author} {\bibfnamefont {L.}~\bibnamefont {Bottou}},
  \bibinfo {author} {\bibfnamefont {Y.}~\bibnamefont {Bengio}}, \ and\ \bibinfo
  {author} {\bibfnamefont {P.}~\bibnamefont {Haffner}},\ }\href@noop {}
  {\bibfield  {journal} {\bibinfo  {journal} {Proceedings of the IEEE}\
  }\textbf {\bibinfo {volume} {86}},\ \bibinfo {pages} {2278} (\bibinfo {year}
  {1998})}\BibitemShut {NoStop}%
\bibitem [{\citenamefont {Albawi}\ \emph {et~al.}(2017)\citenamefont {Albawi},
  \citenamefont {Mohammed},\ and\ \citenamefont
  {Al-Zawi}}]{albawi2017understanding}%
  \BibitemOpen
  \bibfield  {author} {\bibinfo {author} {\bibfnamefont {S.}~\bibnamefont
  {Albawi}}, \bibinfo {author} {\bibfnamefont {T.~A.}\ \bibnamefont
  {Mohammed}}, \ and\ \bibinfo {author} {\bibfnamefont {S.}~\bibnamefont
  {Al-Zawi}},\ }in\ \href@noop {} {\emph {\bibinfo {booktitle} {2017
  International Conference on Engineering and Technology (ICET)}}}\ (\bibinfo
  {organization} {Ieee},\ \bibinfo {year} {2017})\ pp.\ \bibinfo {pages}
  {1--6}\BibitemShut {NoStop}%
\bibitem [{\citenamefont {Lu}\ \emph {et~al.}(2018)\citenamefont {Lu},
  \citenamefont {Huang}, \citenamefont {Li}, \citenamefont {Li}, \citenamefont
  {Chen}, \citenamefont {Lu}, \citenamefont {Ji}, \citenamefont {Shen},
  \citenamefont {Zhou},\ and\ \citenamefont {Zeng}}]{lu2018separability}%
  \BibitemOpen
  \bibfield  {author} {\bibinfo {author} {\bibfnamefont {S.}~\bibnamefont
  {Lu}}, \bibinfo {author} {\bibfnamefont {S.}~\bibnamefont {Huang}}, \bibinfo
  {author} {\bibfnamefont {K.}~\bibnamefont {Li}}, \bibinfo {author}
  {\bibfnamefont {J.}~\bibnamefont {Li}}, \bibinfo {author} {\bibfnamefont
  {J.}~\bibnamefont {Chen}}, \bibinfo {author} {\bibfnamefont {D.}~\bibnamefont
  {Lu}}, \bibinfo {author} {\bibfnamefont {Z.}~\bibnamefont {Ji}}, \bibinfo
  {author} {\bibfnamefont {Y.}~\bibnamefont {Shen}}, \bibinfo {author}
  {\bibfnamefont {D.}~\bibnamefont {Zhou}}, \ and\ \bibinfo {author}
  {\bibfnamefont {B.}~\bibnamefont {Zeng}},\ }\href@noop {} {\bibfield
  {journal} {\bibinfo  {journal} {Physical Review A}\ }\textbf {\bibinfo
  {volume} {98}},\ \bibinfo {pages} {012315} (\bibinfo {year}
  {2018})}\BibitemShut {NoStop}%
\bibitem [{\citenamefont {Doherty}\ \emph {et~al.}(2002)\citenamefont
  {Doherty}, \citenamefont {Parrilo},\ and\ \citenamefont
  {Spedalieri}}]{doherty2002distinguishing}%
  \BibitemOpen
  \bibfield  {author} {\bibinfo {author} {\bibfnamefont {A.~C.}\ \bibnamefont
  {Doherty}}, \bibinfo {author} {\bibfnamefont {P.~A.}\ \bibnamefont
  {Parrilo}}, \ and\ \bibinfo {author} {\bibfnamefont {F.~M.}\ \bibnamefont
  {Spedalieri}},\ }\href@noop {} {\bibfield  {journal} {\bibinfo  {journal}
  {Physical Review Letters}\ }\textbf {\bibinfo {volume} {88}},\ \bibinfo
  {pages} {187904} (\bibinfo {year} {2002})}\BibitemShut {NoStop}%
\bibitem [{\citenamefont {Wei}(2008)}]{wei2008relative}%
  \BibitemOpen
  \bibfield  {author} {\bibinfo {author} {\bibfnamefont {T.-C.}\ \bibnamefont
  {Wei}},\ }\href@noop {} {\bibfield  {journal} {\bibinfo  {journal} {Physical
  Review A}\ }\textbf {\bibinfo {volume} {78}},\ \bibinfo {pages} {012327}
  (\bibinfo {year} {2008})}\BibitemShut {NoStop}%
\bibitem [{\citenamefont {Peruzzo}\ \emph {et~al.}(2014)\citenamefont
  {Peruzzo}, \citenamefont {McClean}, \citenamefont {Shadbolt}, \citenamefont
  {Yung}, \citenamefont {Zhou}, \citenamefont {Love}, \citenamefont
  {Aspuru-Guzik},\ and\ \citenamefont {O’brien}}]{peruzzo2014variational}%
  \BibitemOpen
  \bibfield  {author} {\bibinfo {author} {\bibfnamefont {A.}~\bibnamefont
  {Peruzzo}}, \bibinfo {author} {\bibfnamefont {J.}~\bibnamefont {McClean}},
  \bibinfo {author} {\bibfnamefont {P.}~\bibnamefont {Shadbolt}}, \bibinfo
  {author} {\bibfnamefont {M.-H.}\ \bibnamefont {Yung}}, \bibinfo {author}
  {\bibfnamefont {X.-Q.}\ \bibnamefont {Zhou}}, \bibinfo {author}
  {\bibfnamefont {P.~J.}\ \bibnamefont {Love}}, \bibinfo {author}
  {\bibfnamefont {A.}~\bibnamefont {Aspuru-Guzik}}, \ and\ \bibinfo {author}
  {\bibfnamefont {J.~L.}\ \bibnamefont {O’brien}},\ }\href@noop {} {\bibfield
   {journal} {\bibinfo  {journal} {Nature communications}\ }\textbf {\bibinfo
  {volume} {5}},\ \bibinfo {pages} {1} (\bibinfo {year} {2014})}\BibitemShut
  {NoStop}%
\bibitem [{\citenamefont {Mitarai}\ \emph {et~al.}(2018)\citenamefont
  {Mitarai}, \citenamefont {Negoro}, \citenamefont {Kitagawa},\ and\
  \citenamefont {Fujii}}]{mitarai2018quantum}%
  \BibitemOpen
  \bibfield  {author} {\bibinfo {author} {\bibfnamefont {K.}~\bibnamefont
  {Mitarai}}, \bibinfo {author} {\bibfnamefont {M.}~\bibnamefont {Negoro}},
  \bibinfo {author} {\bibfnamefont {M.}~\bibnamefont {Kitagawa}}, \ and\
  \bibinfo {author} {\bibfnamefont {K.}~\bibnamefont {Fujii}},\ }\href@noop {}
  {\bibfield  {journal} {\bibinfo  {journal} {Physical Review A}\ }\textbf
  {\bibinfo {volume} {98}},\ \bibinfo {pages} {032309} (\bibinfo {year}
  {2018})}\BibitemShut {NoStop}%
\bibitem [{\citenamefont {McClean}\ \emph {et~al.}(2016)\citenamefont
  {McClean}, \citenamefont {Romero}, \citenamefont {Babbush},\ and\
  \citenamefont {Aspuru-Guzik}}]{mcclean2016theory}%
  \BibitemOpen
  \bibfield  {author} {\bibinfo {author} {\bibfnamefont {J.~R.}\ \bibnamefont
  {McClean}}, \bibinfo {author} {\bibfnamefont {J.}~\bibnamefont {Romero}},
  \bibinfo {author} {\bibfnamefont {R.}~\bibnamefont {Babbush}}, \ and\
  \bibinfo {author} {\bibfnamefont {A.}~\bibnamefont {Aspuru-Guzik}},\
  }\href@noop {} {\bibfield  {journal} {\bibinfo  {journal} {New Journal of
  Physics}\ }\textbf {\bibinfo {volume} {18}},\ \bibinfo {pages} {023023}
  (\bibinfo {year} {2016})}\BibitemShut {NoStop}%
\bibitem [{\citenamefont {Mari}\ \emph {et~al.}(2020)\citenamefont {Mari},
  \citenamefont {Bromley}, \citenamefont {Izaac}, \citenamefont {Schuld},\ and\
  \citenamefont {Killoran}}]{mari2020transfer}%
  \BibitemOpen
  \bibfield  {author} {\bibinfo {author} {\bibfnamefont {A.}~\bibnamefont
  {Mari}}, \bibinfo {author} {\bibfnamefont {T.~R.}\ \bibnamefont {Bromley}},
  \bibinfo {author} {\bibfnamefont {J.}~\bibnamefont {Izaac}}, \bibinfo
  {author} {\bibfnamefont {M.}~\bibnamefont {Schuld}}, \ and\ \bibinfo {author}
  {\bibfnamefont {N.}~\bibnamefont {Killoran}},\ }\href@noop {} {\bibfield
  {journal} {\bibinfo  {journal} {Quantum}\ }\textbf {\bibinfo {volume} {4}},\
  \bibinfo {pages} {340} (\bibinfo {year} {2020})}\BibitemShut {NoStop}%
\bibitem [{\citenamefont {Liu}\ \emph {et~al.}(2021)\citenamefont {Liu},
  \citenamefont {Lim}, \citenamefont {Wood}, \citenamefont {Huang},
  \citenamefont {Guo},\ and\ \citenamefont {Huang}}]{liu2021hybrid}%
  \BibitemOpen
  \bibfield  {author} {\bibinfo {author} {\bibfnamefont {J.}~\bibnamefont
  {Liu}}, \bibinfo {author} {\bibfnamefont {K.~H.}\ \bibnamefont {Lim}},
  \bibinfo {author} {\bibfnamefont {K.~L.}\ \bibnamefont {Wood}}, \bibinfo
  {author} {\bibfnamefont {W.}~\bibnamefont {Huang}}, \bibinfo {author}
  {\bibfnamefont {C.}~\bibnamefont {Guo}}, \ and\ \bibinfo {author}
  {\bibfnamefont {H.-L.}\ \bibnamefont {Huang}},\ }\href@noop {} {\bibfield
  {journal} {\bibinfo  {journal} {Science China Physics, Mechanics \&
  Astronomy}\ }\textbf {\bibinfo {volume} {64}},\ \bibinfo {pages} {1}
  (\bibinfo {year} {2021})}\BibitemShut {NoStop}%
\end{thebibliography}%

\end{document}